\documentclass{Interspeech}

\interspeechcameraready

\title{Leveraging Large Language Models for Spontaneous Speech-Based Suicide Risk Detection}

\author[affiliation={1,2}]{Yifan}{Gao}
\author[affiliation={1,2}]{Jiao}{Fu}
\author[affiliation={1,2}]{Long}{Guo}
\author[affiliation={1,2,*}]{Hong}{Liu}
\affiliation{College of Computer Science}{Sichuan University}{China}
\affiliation{National Key Laboratory of Fundamental Science on Synthetic Vision}{Sichuan University}{China}

\email{\{gaoyifan1,2024226040006,2022223040087\}@stu.scu.edu.cn,\{liuhong\}@scu.edu.cn}

\keywords{Suicide Risk Detection, Large Language Model, Multimodal}

\usepackage{comment}

\usepackage{inconsolata}
\usepackage{float}
\renewcommand{\arraystretch}{1.65} % 可以根据需要修改倍数
\usepackage{amsmath}
\usepackage{multirow}
\usepackage{hyperref}
\usepackage{microtype}
\usepackage[margin=1in]{geometry}

\begin{document}

\maketitle


\begin{abstract}
\renewcommand\thefootnote{}\footnotetext{* Corresponding author}
% 自杀风险的早期识别对于预防自杀行为至关重要，识别并研究与自杀风险相关的模式与标记也因此成为当前研究的重点。在本文中，我们展示了在 The 1st SpeechWellness Challenge 中的尝试结果。该挑战赛旨在探索言语作为一种非侵入性且易获取的心理健康指标，用于识别具有自杀风险的青少年。
% 我们的方法以大语言模型作为主要的特征提取工具，结合常规的声学特征与语义特征，在测试集上实现了 74% 的准确率，在SW1挑战赛中名列第一。我们的结果表明，大语言模型在自杀风险言语分析中具有新颖且有影响力的应用价值
Early identification of suicide risk is crucial for preventing suicidal behaviors. As a result, the identification and study of patterns and markers related to suicide risk have become a key focus of current research. In this paper, we present the results of our work in the 1st SpeechWellness Challenge (SW1), which aims to explore speech as a non-invasive and easily accessible mental health indicator for identifying adolescents at risk of suicide.
Our approach leverages large language model (LLM) as the primary tool for feature extraction, alongside conventional acoustic and semantic features. The proposed method achieves an accuracy of 74\% on the test set, ranking first in the SW1 challenge. These findings demonstrate the potential of LLM-based methods for analyzing speech in the context of suicide risk assessment.

\end{abstract}

\section{Introduction}

Suicide is a significant global mental health issue, with over 720,000 deaths annually and the number of people who attempt suicide is much larger than that of those who die by suicide. 
This mental health issue is particularly prominent among young people, as suicide ranks as the third leading cause of death globally for individuals aged 15 to 29 \cite{WHO2024Suicide}.
Individuals with suicidal tendencies do not exhibit singular or distinct clinical characteristics, making the identification of suicidal ideation an exceptionally challenging task. 
% 自杀是一个严重的全球心理健康问题，每年有超过 72 万人因此死亡，且尝试自杀的人数远远超过真正因自杀而死亡的人数。
% 这一心理健康问题在年轻人群体中尤为突出：在全球 15 至 29 岁的人群中，自杀是第三大死因 \cite{WHO2024Suicide}。
% 具有自杀倾向的个体并不呈现单一或明显的临床特征，这使得识别自杀意念成为一项极具挑战性的任务。

Traditional suicide assessment methods primarily include clinical interviews \cite{talk} and psychological measurement scales \cite{paper}. 
However, the accuracy of these methods can be compromised by the patient's state of mind when conveying their symptoms, emotions, or cognitive experiences, potentially leading to discrepancies between the assessment results and the actual situation \cite{cummins2015review}.
This is because traditional assessment methods heavily rely on patients' voluntary disclosure, and many individuals with suicidal ideation struggle to express their true thoughts candidly during interviews due to psychological barriers \cite{ballard2021new}.
This has driven research into automatic suicide risk detection.
% 传统的自杀评估方法主要包括临床访谈 \cite{talk} 和心理测量量表 \cite{paper}。
% 然而，由于患者在表达自身症状、情绪或认知体验时所处的心理状态可能会影响结果，这些方法的准确性往往受到削弱，导致评估结果与实际情况之间可能出现差异 \cite{cummins2015review}。
% 这是因为传统的评估方法在很大程度上依赖患者主动披露信息，而许多有自杀意念的个体由于心理障碍，在访谈过程中难以坦诚地表达真实想法 \cite{ballard2021new}。
% 正因如此，研究人员开始探索自动化自杀风险检测。

Speech, as a signal rich in information and a quantifiable behavior, can be collected outside clinical settings. This characteristic of speech enhances patients'access to care and enables real-time, context-aware monitoring of an individual's mental state \cite{cummins2015review,johar2015emotion}.
%语音作为一种蕴含丰富信息的信号与可量化的行为，可以在临床环境之外收集，它的这种特点可以增加患者获得护理的机会，并实现对个人精神状态的实时和情境感知监测[13、14]。
An increasing body of research indicates that speech analysis based on semantic textual features plays a significant role in screening and identifying patients at risk of suicide \cite{belouali2021acoustic}.
%越来越多的研究表明，基于语义文本特征的语音分析在筛查和识别有自杀风险的患者方面发挥着重要作用。
Individuals with a history of suicidal behavior or suicidal ideation often show distinct differences in word choice \cite{HOMAN2022102161}and narrative perspective \cite{bs14030225} when reading aloud or engaging in conversation, compared to healthy controls. 
%与健康对照组相比，有自杀行为或自杀意念史的个体在大声朗读或参与对话时，在词汇选择和叙事视角上往往表现出明显的差异。 
These individuals typically face greater levels of stress, are frequently subjected to blame \cite{Hagan03072017}in their daily lives, and lack effective means to cope with that stress \cite{HORWITZ20111077}, all of which heighten their emotional and psychological burdens. 
%这些人通常面临更大程度的压力，在日常生活中经常受到指责，并且缺乏应对压力的有效手段，所有这些都会增加他们的情感和心理负担。
Studies utilizing large-scale social media \cite{media,media1} have demonstrated that these semantic textual features offer a highly promising avenue for the real-time screening of high-risk populations.
%大规模流行病学研究表明，这些语义文本特征为实时筛查高危人群提供了非常有希望的途径。
For widespread application, it is essential to develop fully automated, reliable, and interpretable systems for real-time prediction and clinical intervention.
%为了使这些方法得到广泛应用，必须开发能够支持实时预测和临床干预的全自动、可靠和可解释的系统。

\begin{figure*}[htbp]
  \centering
  \includegraphics[width=\textwidth]{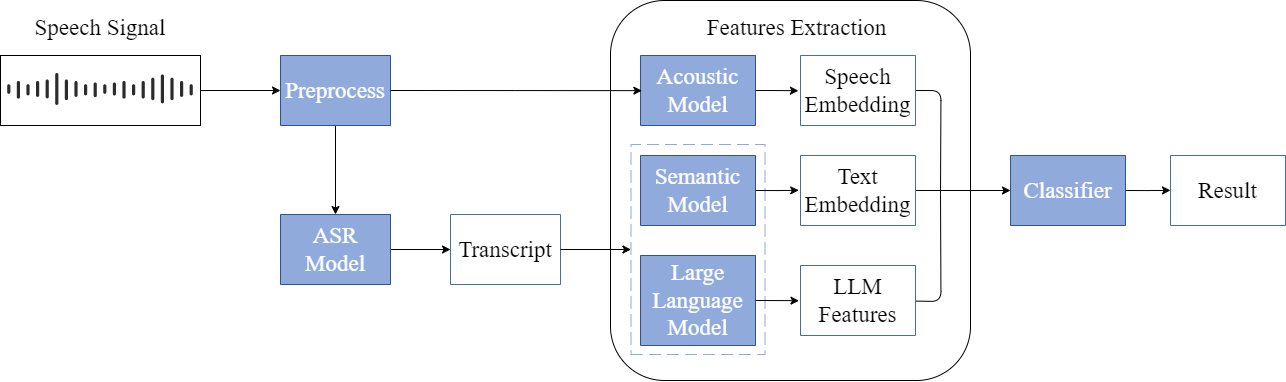}
  \caption{The framework of the proposed method.}
  \label{fig:pipeline}
\end{figure*}

The emergence of large language model (LLM) has introduced new possibilities for automated natural language processing. While LLM have shown promising outcomes in speech detection for other diseases such as depression \cite{depress} and Alzheimer's disease \cite{AD}, their application to suicide risk analysis remains comparatively underdeveloped. Most existing methods still heavily rely on traditional acoustic and linguistic features, or legacy language models like BERT, for risk prediction \cite{ananthakrishnan2022suicidal, metzler2022detecting}. \cite{cui2024spontaneous} demonstrated that large language models can achieve favorable results for suicide risk detection; however, their approach lacked interpretability, limiting its usefulness in contexts that require transparent decision-making and clinical intervention.
% 近年来，大语言模型（LLMs）的兴起为自动化自然语言处理开辟了新的可能性。尽管 LLMs 已被广泛应用于其他疾病的语音检测之中,比如抑郁症\cite{depress}，阿尔兹海默症\cite{AD}，取得了很好的效果
%但在自杀风险的语音分析方面，相较之下仍处于相对不发达的阶段。
%当前的主流方法仍主要依赖于传统的（声学和语言）特征，或是使用较早的语言模型（如 BERT）进行风险预测\cite{ananthakrishnan2022suicidal,devika20221d,metzler2022detecting}。
% \cite{cui2024spontaneous} 探索了使用大语言模型进行自杀风险预测，并取得了有希望的结果；然而，他们的方法缺乏可解释性，而在需要透明决策过程和临床干预的场景中，这往往会限制其实用性。

Finally, studies show that integrating multiple data modalities can boost mental health diagnostics. For example, combining speech, text, and physiological signals for depression detection often outperforms single-modality approaches \cite{zhang2024multimodal}. These methods capture complementary features, providing a more holistic view of an individual’s mental health status.
% 最后，越来越多的研究证据表明，融合多模态数据能够显著提升心理健康诊断系统的性能。例如，在阿尔茨海默病或抑郁症检测中，将语音（声学）、文本（语言学）乃至生理信号等多种模态进行结合，相比仅使用单一模态，往往能够取得更优的表现 \cite{zhang2024multimodalD}。这些多模态方法可以提取相对独立且互补的特征，从而为个体的心理健康状况提供更全面、细致的表征。

Building on previous work, this study focuses on the early identification of suicide risk to prevent suicidal behaviors. To achieve this, we propose a novel method for speech-based suicide risk detection, leveraging LLM as the primary tool for feature extraction, alongside conventional acoustic and semantic features.
% 在已有研究的基础上，本研究聚焦于自杀风险的早期识别，以预防自杀行为。为此，我们提出了一种基于语音进行自杀风险检测的新方法，利用大语言模型（LLM）作为主要特征提取工具，并结合传统声学与语义特征。
Our major contributions are listed below:

\begin{itemize}
    \item We utilize LLM to automatically extract interpretable features related to suicide risk from spontaneous speech transcripts. By employing structured prompts and standardized evaluation procedures, we enhance the transparency and interpretability of semantic indicators.
    \item We further integrate conventional acoustic and semantic features with the extracted LLM features, forming a multimodal framework that ensures more robust suicide risk detection.
    \item Our method achieve 74\% accuracy on the challenge dataset, ranking first in the challenge, thereby demonstrating the effectiveness and potential applications of our approach in suicide risk detection.

\end{itemize}

% 我们基于大语言模型（LLM）自动从自发语音转录文本中提取与自杀风险相关的可解释特征。通过结构化提示和标准化评估流程，提升了语义指标的透明度与可解释性。

% 我们在提取的 LLM 特征基础上，进一步融合常规声学和语义特征，构建了一个多模态检测框架，以实现更加稳健的自杀风险识别。

% 我们的方法在本次挑战赛提供的数据集上取得了 74% 的准确率，排名第一，充分证明了该方法在自杀风险分析中的有效性与潜在应用价值

\section{Dataset}

As shown in Table \ref{tab:dataset}. The SW1 dataset \cite{baseline} consists of a training set and a development set (dev).
The training set contains data from 400 participants, while the development set includes data from 100.
The data distribution in both datasets is the same, with 50\% of the participants diagnosed as having a suicide risk.
The recordings used for the SW1 challenge consist of each participant’s responses to three different speech tasks designed to capture a range of linguistic, cognitive, and emotional features:

Emotional Regulation (ER): Participants are asked to respond to an open-ended question: “Have you ever experienced moments of extreme emotional distress? How do you manage these feelings ?”

Passage Reading (PR): Participants read the fable “The North Wind and the Sun” from Aesop’s Fables, commonly used in linguistic studies for analyzing structures such as morphemes.

Expression Description (ED): Given an image of a facial expression, participants are asked to describe it.

% 如表 \ref{tab:dataset} 所示，SW1 数据集 \cite{baseline} 由训练集和开发集（dev）组成。训练集包含来自 400 名受试者的数据，开发集则包含来自 100 名受试者的数据。两部分数据的分布相同，均有 50% 的受试者被诊断为存在自杀风险。
% 在 SW1 挑战赛中所使用的录音包含每位受试者对三种不同语音任务的回答，这些任务旨在提取多样的语言、认知与情感特征：

% 情绪调节（ER）：受试者需要回答一个开放式问题，“你曾经经历过极度情绪困扰吗？你是如何应对这些情绪的？”
% 短文朗读（PR）：受试者朗读诗歌《北风与太阳》，该文本出自伊索寓言，常用于语言学研究（如考察语素结构等）。
% 表情描述（ED）：给出一张面部表情的图片，请受试者对其进行描述。

{
\renewcommand{\arraystretch}{1.3} % 在此处局部恢复为原先值或使用其他值
\begin{table}[htbp]
    \centering
    % 优化表格标题，设置字体大小和样式
    \captionsetup{font=small,labelfont=bf}
    \caption{The gender ratio and average age of young people with and without suicidal tendencies}
    \label{tab:dataset}
    \begin{tabular}{cccc}
        \toprule
        % 使用 \Centering 让列标题居中
        Suicide Risk &Gender (M:F) &Average Age \\
        \midrule
        Yes  & 61:189 & 14.06   \\
        No  & 93:157 & 13.48     \\
        \bottomrule
    \end{tabular}
\end{table}
}

\section{Method}

As illustrated in Figure \ref{fig:pipeline}, we process raw audio in two main steps.
First, the audio is fed into an acoustic model to extract acoustic features, while an Automatic Speech Recognition (ASR) model transcribes the speech into text.
Next, the transcription is input into a semantic model to extract text embeddings and into an LLM for LLM features.
Finally, each set of modality-specific features is classified, and the outputs are combined to produce the final prediction.
% 如图 \ref{fig:pipeline} 所示，我们对原始音频的处理分为两个主要步骤。
% 首先，将音频输入声学模型以提取声学特征，同时由自动语音识别（ASR）模型将语音转录为文本。
% 接着，将转录得到的文本输入语义模型以提取文本嵌入，并输入大型语言模型（LLM）以获取 LLM 特征。
% 最后，对每一种特定模态的特征分别进行分类，并将这些输出结果组合，生成最终预测。

\subsection{Preprocess}

To further enhance the model’s performance,  we apply Koala\footnote{\href{https://picovoice.ai/docs/api/koala-web/}{https://picovoice.ai/docs/api/koala-web}} for noise reduction on the recordings in our dataset. 
Koala combines time-frequency analysis with a deep neural network to effectively separate background noise from speech, thereby improving clarity and reducing environmental interference.

After noise reduction, we segment the processed audio into 30-second intervals with about 10\% overlap between adjacent segments, aiming to minimize the truncation of important features at segment boundaries and to retain transitional information. 
Finally, we perform average pooling on the acoustic features across all segments for each speaker, resulting in a more robust speaker-level representation.

\subsection{Acoustic Model} 

\noindent \textbf{Wav2Vec2-large-xlsr-53.\footnote{\href{https://huggingface.co/facebook/wav2vec2-large-xlsr-53}{https://huggingface.co/facebook/wav2vec2-large-xlsr-53}}}
Wav2Vec2-large-xlsr-53 \cite{w2v} is a multilingual extension of Wav2Vec2, equipped with 24 Transformer layers and a hidden size of 1024.
It is pre-trained on approximately 56,000 hours of unlabeled speech data covering 53 languages. 
% This model utilizes self-supervised contrastive learning by masking parts of the audio features and predicting the latent representations, allowing it to learn robust and universal speech features.

\noindent \textbf{Whisper-large-v3.\footnote{\href{https://huggingface.co/openai/whisper-large-v3}{https://huggingface.co/openai/whisper-large-v3}}}
Whisper-large-v3 \cite{whisper} is the large version of OpenAI Whisper, featuring a Transformer architecture with 32 encoder layers and 32 decoder layers, along with a hidden size of 1280. 
It is pre-trained on 680,000 hours of multilingual audio data. 
% The model leverages an encoder-decoder Transformer architecture to perform end-to-end sequence-to-sequence transformation, enabling multilingual speech recognition and translation.

\noindent \textbf{Hubert-large-ls960-ft.\footnote{\href{https://huggingface.co/facebook/hubert-large-ls960-ft}{https://huggingface.co/facebook/hubert-large-ls960-ft}}}
Hubert-large-ls960-ft \cite{hubert} is a large version of Hubert, featuring 24 Transformer layers and a hidden size of 1024. 
It is first pre-trained on 60,000 hours of unlabeled speech data (Libri-Light), then fine-tuned on 960 hours of labeled LibriSpeech data.

\subsection{Sementic Model}
\textbf{BERT-base-Chinese.\footnote{\href{https://huggingface.co/google-bert/bert-base-chinese}{https://huggingface.co/google-bert/bert-base-chinese}}}
BERT-base-Chinese  \cite{bert} is a large-scale pretrained language model specialized for Chinese, featuring 12 Transformer layers and a hidden size of 768. 
It is trained on extensive Chinese corpora (such as Chinese Wikipedia), where it learns deep bidirectional representations by predicting masked tokens and performing next-sentence classification.

\noindent \textbf{XLM-RoBERTa-base.\footnote{\href{https://huggingface.co/FacebookAI/xlm-roberta-base}{https://huggingface.co/FacebookAI/xlm-roberta-base}}}
XLM-RoBERTa-base \cite{robert} is a multilingual pretrained language model by Facebook AI that supports over 100 languages, including both high- and low-resource ones. 
It shares RoBERTa's architecture with 12 Transformer layers and a hidden size of 768. Trained on large-scale cross-lingual corpora using self-supervised tasks (e.g., masked language modeling), it learns universal cross-lingual representations.

\subsection{Large Language Model}

\textbf{DeepSeek-R1.\footnote{\href{https://huggingface.co/deepseek-ai/DeepSeek-R1}{https://huggingface.co/deepseek-ai/DeepSeek-R1}}} 
DeepSeek-R1 \cite{deepseek} is one of the most popular and powerful LLMs for reasoning.
It is trained using reinforcement learning and demonstrates strong capabilities in mathematics, coding, and natural language reasoning. 
It excels in multiple benchmark tests and can autonomously verify and reflect, enhancing the accuracy and coherence of its reasoning

We programmatically extract relevant features from the transcribed text with the DeepSeek-R1 large language model.
To identify these features preliminarily, we perform basic preprocessing on the transcriptions of the Emotional Regulation (ER) task audio recordings, such as removing extra spaces and normalizing symbols. 
Subsequently, we use DeepSeek-R1 to check for expressions in the text related to sharing emotions, signs of avoiding emotions, using exercise to relieve stress, and references to self-harm behaviors.

Previous research has shown that social support \cite{wickramaratne2022social,social1,social2} and positive exercise habits \cite{sports1,sports2} can significantly reduce suicide risk, while negative stress-relief behaviors and conflicts with parents or friends may increase the risk \cite{outlet1,outlet2}.
DeepSeek-R1 successfully captures interpretable features related to emotional dysregulation, enhancing the model's accuracy and interpretability in future emotional regulation applications, thus advancing the research.

For each feature, we query the DeepSeek-R1 model to check if similar behaviors are present, marking them as 1 if present and 0 if not. 
In addition to numeric evaluation, we also ask DeepSeek-R1 to extract the corresponding text segments from the transcription to explain the evaluation results.
The original transcription text is in Chinese, and both the prompts and example outputs have been translated, as shown in Figure \ref{fig:deepseek-prompt}.

% DeepSeek-R1\cite{deepseek} 是目前最受欢迎且功能最强大的推理型大型语言模型（LLM）之一。
% 它通过强化学习进行训练，在数学、编程以及自然语言推理等方面展现了强大的能力。
% 在多个基准测试（benchmark）中表现出色，并且能够自主验证与反思，从而提高推理的准确性和连贯性。

% 在本研究中，我们使用 DeepSeek-R1 大型语言模型对转录文本进行程序化地特征提取。
% 为初步识别这些特征，我们首先对情绪调节（Emotional Regulation, ER）任务音频录音的转录文本进行基本预处理，包括去除多余空格和规范化符号。
% 随后，我们利用 DeepSeek-R1 对文本中是否存在与向他人分享情感、回避情感、通过运动缓解压力以及是否提及自残行为等相关的表达进行检查。

% 先前的研究表明，社会支持\cite{wickramaratne2022social,social1,social2}和积极的运动方式\cite{sports1,sports2}都可以显著降低自杀风险，而消极的压力纾解行为则可能增加风险\cite{outlet1,outlet2}。
% DeepSeek-R1 成功捕捉到了与情绪失调相关的可解释特征，这有助于在未来的情绪调节应用中提高模型的准确性与可解释性，从而推动本研究的进展。

% 针对每个特征，我们向 DeepSeek-R1 模型查询是否存在类似的行为，如果存在则记为 1，否则记为 0。
% 除了数字化的评估结果，我们还要求 DeepSeek-R1 从转录文本中提取相应的文本片段，用以解释评估结果。
% 由于原始转录文本是中文，提示词和示例输出经过翻译，如图 \ref{fig:deepseek-prompt} 所示。

%--------------------------------------------以下是对figure的解释，非正文------------------------------

\begin{figure}[!htbp]

    \centering
    \scalebox{0.75}{
        \begin{minipage}{10.5cm}
            \textbf{Prompt:\quad}\texttt{Here are key indicators of suicide risk in spontaneous speech:} \\\\
            \texttt{
            Self-harm Behavior: Self-harm Behavior refers to any explicit mention or description of actions intended to harm oneself, such as cutting, overdosing, hitting one’s head against a wall, or using any kind of implement to inflict injury. This can be either a direct reference or an indirect hint that indicates deliberately causing physical harm to the body. Any indication of self-inflicted pain, whether recent or in the past, would fall under this category.\\\\
            Pressure: Pressure primarily arises from conflicts or tense interactions with parents and friends, leading to emotional distress or anxiety. Situations such as being harshly criticized by family members or having disputes with peers can significantly heighten stress levels, affecting an individual’s mood and overall mental well-being. By focusing on these direct interpersonal triggers, it becomes clearer how specific sources of pressure may contribute to negative emotional experiences.\\\\
            Social Support: Social Support focuses on whether the interviewee seeks or receives help from family, friends, mentors, or professionals when experiencing negative emotions or distress. The key is whether they share their feelings with others or rely on the presence or advice of someone close to them. Mentioning any form of talking to parents, friends, or a counselor about worries or sadness signals that they do not face stress entirely alone, which may lower the level of risk.\\\\
            Unhealthy Outlets: Unhealthy Outlets refer to harmful or isolating methods of coping with negative emotions or pressure. Destructive actions such as breaking objects, physically lashing out, or causing harm indicate an inability to manage stress in a constructive way. Likewise, retreating from others, refusing communication, or locking oneself away can also signify an unhealthy approach to emotional struggles. These behaviors neither resolve the underlying issues nor foster healthier coping mechanisms, potentially increasing overall risk.\\\\
            Exercise: Exercise here specifically refers to structured or recognized forms of physical activity undertaken to alleviate tension and improve mood. Sports and workouts such as running, swimming, playing basketball, lifting weights, or engaging in any organized fitness routine count as exercise in this context. Simple walks, breathwork, stretching, or casual movements do not fulfill this indicator, as the focus is on more vigorous or intentional physical training.\\\\
            For each of these indicators, and for the following transcript, please indicate whether each indicator is fulfilled (0 = no, 1 = yes). Then provide 1–3 examples from the text for your assessment (in brackets, marking verbatim quotations with "). Follow these instructions carefully:\\
            1. Only use the English translations in brackets, ensuring that the bracketed content is in English (short verbatim quotes).\\
            2.If an indicator is fulfilled (1 = yes), provide the bracketed quotes. If an indicator is not fulfilled (0 = no), do not provide any bracketed text.\\
            3.Provide nothing else and no explanation. Do not add any commentary, summary, or alternative phrasing outside of the bracketed quotes.\\
            4.Retain punctuation in the bracketed English quotes as accurately as possible.\\
            Note: The interviewee’s speech is in Chinese, and was transcribed and analyzed in Chinese. The bracketed content is an English translation of the original Chinese text, not the original text itself.\\
            \\
            \{transcript\}} \\\\
            \textbf{DeepSeek-R1 response:}
            \texttt{\\
             Self-harm Behavior: 0 \\
             Pressure: 1 ["my parents say I’m not good enough","they say some very awful things"]\\
             Social Support: 1 ["complained to my family", "felt calm after talking with my family"]\\
             Unhealthy Outlets: 0 \\
             Exercise: 1 ["exercised at home on the weekend"]\\
          }
            
        \end{minipage}
           }
        \caption {An Example of Using Prompts to Extract DeepSeek Suicide Risk Indicators}
    \label{fig:deepseek-prompt}
\end{figure}

\subsection{Classification}

In this study, we employ distinct classification pipelines for suicide risk detection using acoustic features, textual features, and interpretable features. First, acoustic embeddings extracted from Hubert, Wav2Vec2, or Whisper are fed into a two-layer Multilayer Perceptron (MLP) classifier.
Next, textual embeddings obtained from BERT or RoBERTa are similarly processed by the same two-layer MLP architecture.
Finally, for interpretable features automatically identified by DeepSeek-R1—such as self-harm behavior and social support—we use a Random Forest (RF) classifier.
By pairing each modality with a suitable classification model, our approach effectively exploits the complementary strengths of acoustic signals, textual semantics, and interpretable factors.

\section{Experiment}

\subsection{Implementation Details}
Our model was implemented in PyTorch 2.0.1 on an L20 GPU. We employed a MLP classifier with two hidden layers (64 and 32 units, respectively) and a dropout rate of 0.2.
We trained the model using Adam (learning rate = 1e-4, batch size = 32) for up to 300 epochs, applying early stopping if the validation metric did not improve after 50 epochs.
The dataset followed its predefined partition, with model selection performed via 10-fold cross-validation and final evaluation on the held-out set.

\subsection{Results}

\begin{table}[ht]
\centering
\caption{Performance of different models on various tasks under 10-fold cross-validation.}
\resizebox{\linewidth}{!}{%
\begin{tabular}{c c c c c c}
\hline
\textbf{} & \textbf{Model} & \textbf{Task} & \textbf{Classifier} & \textbf{Accuracy} & \textbf{F1}\\
\hline
1  & \multirow{3}{*}{Hubert}     & ER & MLP & 0.601 & 0.612 \\
2  &                             & PR & MLP & 0.616 & 0.615 \\
3  &                             & ED & MLP & 0.628 & 0.636 \\
\hline
4  & \multirow{3}{*}{Wav2Vec2}   & ER & MLP & 0.622 & 0.637 \\
5  &                             & PR & MLP & 0.626 & 0.628 \\
6  &                             & ED & MLP & 0.636 & 0.633 \\
\hline
7  & \multirow{3}{*}{Whisper}    & ER & MLP & 0.608 & 0.654  \\
8  &                             & PR & MLP & 0.615 & 0.612  \\
9  &                             & ED & MLP & 0.624 & 0.669  \\
\hline
10 & \multirow{3}{*}{BERT}       & ER & MLP & 0.580 & 0.576 \\
11 &                             & PR & MLP & 0.568 & 0.627 \\
12 &                             & ED & MLP & 0.558 & 0.542 \\
\hline
13 & \multirow{3}{*}{RoBERTa}    & ER & MLP & 0.610 & 0.585 \\
14 &                             & PR & MLP & 0.554 & 0.567 \\
15 &                             & ED & MLP & 0.548 & 0.523  \\
\hline
16 & DeepSeek-R1                 & ER & RF  & \textbf{0.652} & \textbf{0.679}  \\
\hline
\end{tabular}
}
\label{tab:results}
\end{table}

From Table \ref{tab:results}, it is evident that the three pre-trained audio models (Hubert, Wav2Vec2, and Whisper) outperform the text-based models (BERT and RoBERTa) in suicide risk detection tasks (ER, PR, ED), as reflected in both accuracy and F1 scores. 
One main reason is that audio models directly capture emotional cues embedded in speech—such as prosody, intonation, and variations in energy—whereas transcribed text alone often overlooks these nuanced emotional and cognitive signals.

Additionally, DeepSeek-R1 achieves the highest accuracy (0.652) and F1 score (0.679) in the ER task, substantially outperforming the other methods.
By leveraging key semantic information extracted from the LLM along with an interpretability mechanism, researchers can trace and understand the rationale behind risk assessments. 
In a highly sensitive task like suicide risk detection, this level of transparency not only bolsters the accuracy of identification but also provides clearer, more reliable support for subsequent clinical interventions and psychological consultations.

Taking a closer look at individual task performance reveals that audio models excel in ED.
For instance, Whisper achieves an F1 score of 0.669, indicating the effectiveness of audio-based feature extraction in capturing subtle emotional expressions crucial for ED.
By contrast, text-based models perform better in ER, likely because they can detect implicit or metaphorical emotional cues within textual content.
These complementary strengths highlight the potential of integrating both audio and textual data for a more comprehensive approach to suicide risk detection.

% 从表 \ref{tab:results} 可以看出，在自杀风险检测的核心任务（ER、PR、ED）上，三种预训练的音频模型（Hubert、Wav2Vec2 和 Whisper）在准确率和 F1 得分方面均优于基于文本的模型（BERT 和 RoBERTa）。这是因为音频模型可以通过韵律、语调、节奏以及能量波动等声学特征，更直接地捕捉到言语中蕴含的情绪线索，而单纯的文本转写往往难以全面刻画这些细微的情绪和认知信号。

% 此外，DeepSeek-R1 模型通过融合可解释特征而展现了显著优势。它将大语言模型（LLM）提取的关键语义信息与音频特征相结合，在 ER 任务中的准确率（0.652）与 F1 得分（0.679）均高于其他所有模型。这说明对于需要深层次情绪理解的任务，多模态以及可解释策略在自杀风险检测中有着突出的价值，特别是在捕捉个体潜在的情感与心理状态时。

% 从具体任务表现来看，音频模型在 ED（情绪描述）任务上优势明显。例如，Whisper 在该任务的 F1 值达到了 0.669，突显了音频特征在挖掘情绪表现方面的效率和精度；而文本模型则在 ER（情绪识别）任务中表现更为突出，因为文本模型更擅长捕捉字里行间隐含或具象的情绪线索，从而在理解并定位情绪触点方面有所优势。通过在这些任务上取得不同维度的性能提升，可以进一步证明在自杀风险检测领域，音频和文本信息各自蕴含独特而互补的价值。

\begin{table}[ht]
\centering
\caption{Experiments with a Voting-Based Approach on the Dev/Test Set}
\label{tab:voting_example}
\resizebox{\linewidth}{!}{%
\begin{tabular}{l c c}
\toprule
\textbf{Model} & \textbf{Accuracy (Dev / Test)} & \textbf{F1 (Dev / Test)} \\
\midrule
Baseline\cite{baseline}                     & 0.530 / 0.510  & -/-         \\
Baseline-bonus\cite{baseline}               & 0.560 / 0.610  & -/-         \\
(1, 2, 5, 6, 8, 9)           & 0.710 / 0.540  & 0.723 / 0.546 \\
(1, 2, 5, 6, 8, 13, 16)      & 0.768 / 0.730  & 0.769 / 0.722 \\
(1, 2, 5, 6, 9, 13, 16)      & 0.770 / \textbf{0.740}  & 0.754 / \textbf{0.740} \\
\bottomrule
\end{tabular}
}
\end{table}

Table \ref{tab:voting_example} shows the integrated voting results derived from the various model index combinations listed in Table \ref{tab:results}. The first voting combination primarily consists of outputs from audio models such as HuBERT, Wav2Vec2, and Whisper, each targeting different tasks (ER, PR, ED). Although this ensemble outperforms the Baseline on the development set, its limited generalization on the test set indicates that relying solely on audio-based features may not fully capture the complexity of suicide risk.

To address this limitation, the second and third voting combinations incorporate RoBERTa (13) and DeepSeek-R1 (16). 
RoBERTa offers text-based insights that complement audio features, while DeepSeek-R1 delivers interpretable key features crucial for identifying core risk factors.
Notably, we increased the voting weight of DeepSeek-R1 to emphasize its importance in capturing critical signals.

By merging acoustic, textual, and interpretable components, the third voting combination achieves an accuracy and F1 score of 0.740 on the test set, substantially surpassing both the Baseline and Baseline-bonus. 
This result underscores the importance of leveraging diversified information sources and highlighting interpretable features in the voting process, thereby enhancing model robustness and predictive performance in this highly sensitive task.

% 表 \ref{tab:voting_example} 展示了从表 \ref{tab:results} 所列多种模型索引组合中，通过投票策略得出的综合结果。第一个投票组合主要由 HuBERT、Wav2Vec2 和 Whisper 等音频模型的输出构成，这些模型分别针对不同任务（ER、PR、ED）。尽管该组合在训练集上优于 Baseline，但在测试集上的泛化能力受到限制，表明仅依赖音频特征难以充分捕捉自杀风险的复杂性。

% 为应对这一局限，第二个和第三个投票组合引入了 RoBERTa (13) 与 DeepSeek-R1 (16)。RoBERTa 能够提供与音频特征互补的文本层面信息，而 DeepSeek-R1 则可输出具有可解释性的关键特征，有助于识别核心风险因素。值得注意的是，我们在投票时提高了 DeepSeek-R1 的权重，以突出其在捕捉关键信号方面的重要作用。

% 通过融合声学、文本和可解释特征，第三个投票组合在测试集上的准确率和 F1 值均达到了 0.740，大幅超过 Baseline 和 Baseline-bonus。这一结果强调了多元信息源的重要性，并通过强化可解释特征在投票过程中的影响，进一步提升了模型在这一高敏感度任务中的稳健性与预测性能。

\section{Conclusion}
In this study, we propose a multimodal suicide risk detection framework for spontaneous speech that integrates acoustic embeddings, text embeddings, and interpretable features extracted by a large language model (LLM).
By combining multiple pretrained models and fusion strategies, our method not only captures essential information—such as emotional and cognitive cues—but also provides interpretable text snippets for each detection decision.
In clinical practice, interpretability is especially crucial, as it helps mental health professionals trace the rationale behind high-risk assessments, thereby enabling more targeted and effective interventions.
Experimental results show that our approach achieves both an accuracy and an F1 score of 0.740 on the test set, significantly outperforming the baseline and further demonstrating the effectiveness of our framework.

It is important to note that the findings of this study are based on the MINI-KID scale, which classifies current suicide risk solely as “at risk” or “no risk,” reflecting participants’ immediate responses at the time of assessment and not serving as a prediction of future suicidal behavior. Although the MINI-KID suicide module is widely recognized as the “gold standard” for evaluating current suicide risk in adolescents, its reliance on self-reported data can lead to underreporting or misinterpretation of symptoms, and its fixed set of items may fail to capture the complex and dynamic nature of suicidal ideation and behavior. Consequently, the results presented here should be interpreted strictly within the context of this particular assessment framework.

% 在本研究中，我们提出了一种基于自发语音的多模态自杀风险检测框架，结合了声学嵌入、文本嵌入以及由大语言模型（LLM）提取的可解释特征。通过多种预训练模型和融合策略的运用，该方法不仅全面捕捉了情绪和认知等关键信息，还为每次检测决策提供了可解释的文本线索。实验结果表明，我们的方法在测试集上取得了 0.740 的准确率和 F1 值，显著优于基准模型，进一步证明了方法的有效性。

% 需要指出的是，本研究的结论基于 MINI-KID 量表，该量表仅通过“有风险”或“无风险”来划分受试者当前的自杀风险，反映的是参与者在评估时的即时反应，并不用于预测未来的自杀行为。尽管 MINI-KID 自杀模块被广泛视为青少年当前自杀风险评估的“黄金标准”，其依赖自我报告的方式仍可能导致漏报或误解，且固定项目难以充分捕捉自杀意念和行为的复杂动态特征。因此，应当在此特定评估框架下谨慎解读本研究所呈现的结果。

\clearpage   
 \section{Acknowledgement}
 This work is partly supported by the Fundamental Research Funds for Central Universities under grants SCU2024D055 and SCU2024D059.

\bibliographystyle{IEEEtran}
\bibliography{mybib}

\begin{thebibliography}{10}
\providecommand{\url}[1]{#1}
\csname url@samestyle\endcsname
\providecommand{\newblock}{\relax}
\providecommand{\bibinfo}[2]{#2}
\providecommand{\BIBentrySTDinterwordspacing}{\spaceskip=0pt\relax}
\providecommand{\BIBentryALTinterwordstretchfactor}{4}
\providecommand{\BIBentryALTinterwordspacing}{\spaceskip=\fontdimen2\font plus
\BIBentryALTinterwordstretchfactor\fontdimen3\font minus \fontdimen4\font\relax}
\providecommand{\BIBforeignlanguage}[2]{{%
\expandafter\ifx\csname l@#1\endcsname\relax
\typeout{** WARNING: IEEEtran.bst: No hyphenation pattern has been}%
\typeout{** loaded for the language `#1'. Using the pattern for}%
\typeout{** the default language instead.}%
\else
\language=\csname l@#1\endcsname
\fi
#2}}
\providecommand{\BIBdecl}{\relax}
\BIBdecl

\bibitem{WHO2024Suicide}
\BIBentryALTinterwordspacing
{World Health Organization}, ``\BIBforeignlanguage{Chinese}{Suicide},'' 2024, accessed: 2024-08-29. [Online]. Available: \url{https://www.who.int/zh/news-room/fact-sheets/detail/suicide}
\BIBentrySTDinterwordspacing

\bibitem{talk}
J.~Oh, K.~Yun, J.-H. Hwang, and J.-H. Chae, ``Classification of suicide attempts through a machine learning algorithm based on multiple systemic psychiatric scales,'' \emph{Frontiers in Psychiatry}, vol.~8, p. 192, 2017.

\bibitem{paper}
D.~J. Dozois and R.~Covin, ``The beck depression inventory-ii (bdi-ii), beck hopelessness scale (bhs), and beck scale for suicide ideation (bss).'' 2004.

\bibitem{cummins2015review}
N.~Cummins, S.~Scherer, J.~Krajewski, S.~Schnieder, J.~Epps, and T.~F. Quatieri, ``A review of depression and suicide risk assessment using speech analysis,'' \emph{Speech communication}, vol.~71, pp. 10--49, 2015.

\bibitem{ballard2021new}
E.~D. Ballard, J.~R. Gilbert, C.~Wusinich, and C.~A. Zarate~Jr, ``New methods for assessing rapid changes in suicide risk,'' \emph{Frontiers in psychiatry}, vol.~12, p. 598434, 2021.

\bibitem{johar2015emotion}
S.~Johar, \emph{Emotion, affect and personality in speech: The Bias of language and paralanguage}.\hskip 1em plus 0.5em minus 0.4em\relax Springer, 2015.

\bibitem{belouali2021acoustic}
A.~Belouali, S.~Gupta, V.~Sourirajan, J.~Yu, N.~Allen, A.~Alaoui, M.~A. Dutton, and M.~J. Reinhard, ``Acoustic and language analysis of speech for suicidal ideation among us veterans,'' \emph{BioData Mining}, vol.~14, no.~1, pp. 1--17, 2021.

\bibitem{HOMAN2022102161}
S.~Homan, M.~Gabi, N.~Klee, S.~Bachmann, A.-M. Moser, M.~Duri', S.~Michel, A.-M. Bertram, A.~Maatz, G.~Seiler, E.~Stark, and B.~Kleim, ``Linguistic features of suicidal thoughts and behaviors: A systematic review,'' \emph{Clinical Psychology Review}, vol.~95, p. 102161, 2022.

\bibitem{bs14030225}
R.~Huang, S.~Yi, J.~Chen, K.~Y. Chan, J.~W.~Y. Chan, N.~Y. Chan, S.~X. Li, Y.~K. Wing, and T.~M.~H. Li, ``Exploring the role of first-person singular pronouns in detecting suicidal ideation: A machine learning analysis of clinical transcripts,'' \emph{Behavioral Sciences}, vol.~14, no.~3, 2024.

\bibitem{Hagan03072017}
C.~R. Hagan and T.~E. Joiner, ``The indirect effect of perceived criticism on suicide ideation and attempts,'' \emph{Archives of Suicide Research}, vol.~21, no.~3, pp. 438--454, 2017, pMID: 27487316.

\bibitem{HORWITZ20111077}
``Specific coping behaviors in relation to adolescent depression and suicidal ideation,'' \emph{Journal of Adolescence}, vol.~34, no.~5, pp. 1077--1085, 2011.

\bibitem{media}
L.~Cao, H.~Zhang, L.~Feng, Z.~Wei, X.~Wang, N.~Li, and X.~He, ``Latent suicide risk detection on microblog via suicide-oriented word embeddings and layered attention,'' \emph{arXiv preprint arXiv:1910.12038}, 2019.

\bibitem{media1}
S.~R. Braithwaite, C.~Giraud-Carrier, J.~West, M.~D. Barnes, and C.~L. Hanson, ``Validating machine learning algorithms for twitter data against established measures of suicidality,'' \emph{JMIR mental health}, vol.~3, no.~2, p. e4822, 2016.

\bibitem{depress}
D.~Shin, H.~Kim, S.~Lee, Y.~Cho, and W.~Jung, ``Using large language models to detect depression from user-generated diary text data as a novel approach in digital mental health screening: Instrument validation study,'' \emph{Journal of Medical Internet Research}, vol.~26, p. e54617, 2024.

\bibitem{AD}
J.~Heitz, G.~Schneider, and N.~Langer, ``Linguistic features extracted by gpt-4 improve alzheimer's disease detection based on spontaneous speech,'' \emph{arXiv preprint arXiv:2412.15772}, 2024.

\bibitem{ananthakrishnan2022suicidal}
G.~Ananthakrishnan, A.~K. Jayaraman, T.~E. Trueman, S.~Mitra, A.~Abinesh, and A.~Murugappan, ``Suicidal intention detection in tweets using bert-based transformers,'' in \emph{2022 International Conference on Computing, Communication, and Intelligent Systems (ICCCIS)}.\hskip 1em plus 0.5em minus 0.4em\relax IEEE, 2022, pp. 322--327.

\bibitem{metzler2022detecting}
H.~Metzler, H.~Baginski, T.~Niederkrotenthaler, and D.~Garcia, ``Detecting potentially harmful and protective suicide-related content on twitter: machine learning approach,'' \emph{Journal of medical internet research}, vol.~24, no.~8, p. e34705, 2022.

\bibitem{cui2024spontaneous}
Z.~Cui, C.~Lei, W.~Wu, Y.~Duan, D.~Qu, J.~Wu, R.~Chen, and C.~Zhang, ``Spontaneous speech-based suicide risk detection using whisper and large language models,'' \emph{arXiv preprint arXiv:2406.03882}, 2024.

\bibitem{zhang2024multimodal}
Z.~Zhang, S.~Zhang, D.~Ni, Z.~Wei, K.~Yang, S.~Jin, G.~Huang, Z.~Liang, L.~Zhang, L.~Li \emph{et~al.}, ``Multimodal sensing for depression risk detection: integrating audio, video, and text data,'' \emph{Sensors}, vol.~24, no.~12, p. 3714, 2024.

\bibitem{baseline}
\BIBentryALTinterwordspacing
W.~Wu, Z.~Cui, C.~Lei, Y.~Duan, D.~Qu, J.~Wu, B.~Zhou, R.~Chen, and C.~Zhang, ``The 1st speechwellness challenge: Detecting suicidal risk among adolescents,'' 2025. [Online]. Available: \url{https://arxiv.org/abs/2501.06474}
\BIBentrySTDinterwordspacing

\bibitem{w2v}
A.~Conneau, A.~Baevski, R.~Collobert, A.~Mohamed, and M.~Auli, ``Unsupervised cross-lingual representation learning for speech recognition,'' \emph{arXiv preprint arXiv:2006.13979}, 2020.

\bibitem{whisper}
A.~Radford, J.~W. Kim, T.~Xu, G.~Brockman, C.~McLeavey, and I.~Sutskever, ``Robust speech recognition via large-scale weak supervision,'' in \emph{International conference on machine learning}.\hskip 1em plus 0.5em minus 0.4em\relax PMLR, 2023, pp. 28\,492--28\,518.

\bibitem{hubert}
W.-N. Hsu, B.~Bolte, Y.-H.~H. Tsai, K.~Lakhotia, R.~Salakhutdinov, and A.~Mohamed, ``Hubert: Self-supervised speech representation learning by masked prediction of hidden units,'' \emph{IEEE/ACM transactions on audio, speech, and language processing}, vol.~29, pp. 3451--3460, 2021.

\bibitem{bert}
J.~Devlin, ``Bert: Pre-training of deep bidirectional transformers for language understanding,'' \emph{arXiv preprint arXiv:1810.04805}, 2018.

\bibitem{robert}
A.~Conneau, ``Unsupervised cross-lingual representation learning at scale,'' \emph{arXiv preprint arXiv:1911.02116}, 2019.

\bibitem{deepseek}
D.~Guo, D.~Yang, H.~Zhang, J.~Song, R.~Zhang, R.~Xu, Q.~Zhu, S.~Ma, P.~Wang, X.~Bi \emph{et~al.}, ``Deepseek-r1: Incentivizing reasoning capability in llms via reinforcement learning,'' \emph{arXiv preprint arXiv:2501.12948}, 2025.

\bibitem{wickramaratne2022social}
P.~J. Wickramaratne, T.~Yangchen, L.~Lepow, B.~G. Patra, B.~Glicksburg, A.~Talati, P.~Adekkanattu, E.~Ryu, J.~M. Biernacka, A.~Charney \emph{et~al.}, ``Social connectedness as a determinant of mental health: A scoping review,'' \emph{PloS one}, vol.~17, no.~10, p. e0275004, 2022.

\bibitem{social1}
A.~M. Brausch and P.~M. Gutierrez, ``Differences in non-suicidal self-injury and suicide attempts in adolescents,'' \emph{Journal of youth and adolescence}, vol.~39, pp. 233--242, 2010.

\bibitem{social2}
E.~M. Kleiman and R.~T. Liu, ``Social support as a protective factor in suicide: Findings from two nationally representative samples,'' \emph{Journal of affective disorders}, vol. 150, no.~2, pp. 540--545, 2013.

\bibitem{sports1}
Y.~Xie, M.~Zhu, X.~Wu, S.~Tao, Y.~Yang, T.~Li, L.~Zou, H.~Xu, and F.~Tao, ``Interaction between physical activity and problematic mobile phone use on suicidality in chinese college students,'' \emph{BMC psychiatry}, vol.~20, pp. 1--7, 2020.

\bibitem{sports2}
M.~Grasdalsmoen, H.~R. Eriksen, K.~J. L{\o}nning, and B.~Sivertsen, ``Physical exercise, mental health problems, and suicide attempts in university students,'' \emph{BMC psychiatry}, vol.~20, pp. 1--11, 2020.

\bibitem{outlet1}
J.~Liang, K.~Kõlves, B.~Lew, D.~de~Leo, L.~Yuan, M.~Abu~Talib, and C.-x. Jia, ``Coping strategies and suicidality: A cross-sectional study from china,'' \emph{Frontiers in Psychiatry}, vol.~11, 2020.

\bibitem{outlet2}
H.~Abdollahpour~Ranjbar, M.~Bakhshesh-Boroujeni, S.~Farajpour-Niri, I.~Hekmati, M.~Habibi~Asgarabad, and M.~Eskin, ``An examination of the mediating role of maladaptive emotion regulation strategies in the complex relationship between interpersonal needs and suicidal behavior,'' \emph{Frontiers in Psychiatry}, vol.~15, 2024.

\end{thebibliography}

\end{document}